\documentclass[%
 aip,
 amsmath,amssymb,
 reprint,%
]{revtex4-1}
\usepackage[normalem]{ulem}
\usepackage{graphicx}
\usepackage{dcolumn}
\usepackage{bm}
\usepackage{xcolor}
\usepackage{soul}

\usepackage[utf8]{inputenc}
\usepackage[T1]{fontenc}
\usepackage{mathptmx}
\usepackage{etoolbox}
\usepackage{physics}
\usepackage{siunitx}
\usepackage{booktabs}
\usepackage[english]{babel} %

\makeatletter
\def\@email#1#2{%
 \endgroup
 \patchcmd{\titleblock@produce}
  {\frontmatter@RRAPformat}
  {\frontmatter@RRAPformat{\produce@RRAP{*#1\href{mailto:#2}{#2}}}\frontmatter@RRAPformat}
  {}{}
}%
\makeatother
\begin{document}

\preprint{AIP/123-QED}
\title{Simulating surfactant effects in phase-transforming fluids}

\author{Keyu Feng}
\affiliation{%
    School of Mechanical Engineering, Purdue University, West Lafayette, IN 47906, USA}
\author{Saikat Mukherjee}
\affiliation{%
    Department of Chemical Engineering, Massachusetts Institute of Technology, Cambridge, MA 02139, USA}
\author{Tianyi Hu}
\affiliation{%
    School of Mechanical Engineering, Purdue University, West Lafayette, IN 47906, USA}
\author{Hector Gomez}
\altaffiliation{Corresponding author: hectorgomez@purdue.edu.}
\affiliation{%
    School of Mechanical Engineering, Purdue University, West Lafayette, IN 47906, USA}

\begin{abstract}
Surfactants are critical in natural processes and engineering, but measuring their concentrations in non-equilibrium conditions and in the presence of flow is difficult. Therefore, computational methods are key tools for improving our understanding. Predicting the effect of surfactants on liquid-vapor transformations is particularly challenging due to (1) simultaneous mass transfer, non-equilibrium thermodynamics and Marangoni stresses, and (2) the phenomenological assumptions underlying many liquid-vapor phase-change models. Starting from the Navier-Stokes-Korteweg equations, a first-principles approach to liquid-vapor phase transformations, we developed a model of liquid-vapor flows with surfactants. We performed simulations of bubbles under equilibrium and liquid-vapor interface oscillations to demonstrate that the model successfully reproduces surfactant-mediated reductions in surface tension. We also investigated the mechanisms whereby surfactant affects bubble coalescence and condensation. Overall, this work provides a new framework for studying the effect of surfactants on liquid-vapor transformations and suggests multiple areas for future research, including the impact of complex surface chemistries on flow around bubbles and the acoustic response of bubbles with surfactants.
\end{abstract}
\maketitle

\section{Introduction}

Surfactants are a general class of chemical compounds that are important in nature and industry due to their ability to reduce the surface tension acting at fluid interfaces. These interfaces may exist between two distinct fluids, such as water and oil; or between the liquid and vapor phases of a single phase-transforming fluid, such as liquid water and water vapor \cite{Rosen2004-qp}. Surfactants are widely present in natural systems and are crucial for vital functions such as promoting pulmonary gas exchange \cite{Mendenhall1964-jm, Longo1993-vh,Dobbs1989-dd,Dahl2005-cm} and improving digestion \cite{Bourbon2001-cj}. Modern science and engineering are utilizing their capacity to alter surface tension, granting them essential roles in detergent formulations \cite{AboulKassim1993-qo}, froth flotation \cite{Pawliszak2024-ut,Pugh1996-zn}, nanomaterial synthesis \cite{Vasudevan2010-em}, pharmaceutical applications \cite{Lawrence1994-qs}, enhanced oil recovery \cite{Massarweh2020-mn}, droplet microfluidics \cite{Baret2012-wh} and waterborne coatings and emulsion polymerization \cite{Lovell2020-co, Pieters2024-rh}.

However, surfactant concentrations may be difficult to measure in heterogeneous environments. In certain cases, we also need spatially resolved concentration measurements because Marangoni stresses depend on surfactant concentration gradients. In the absence of very sophisticated measurement techniques, surfactants act as hidden variables that profoundly impact fluid dynamics. This makes computational models important for quantifying how interfacial flows respond to surfactant distributions \cite{Manikantan2020-nu}. Multicomponent systems with surfactant-covered interfaces have been analyzed using a variety of numerical techniques, including the volume-of-fluid method \cite{Farsoiya2024-io, Alke2009-oc}, level-set formulation \cite{Mialhe2023-bp, Constante-Amores2023-tk, Cleret-de-Langavant2017-go}, phase-field models \cite{Jain2024-ly,Demont2022-nl,Zhu2019-fc, Zhu2019-go, Komura1997-lk}, immersed boundary methods \cite{Khatri2014-ma}, arbitrary Lagrangian-Eulerian methods \cite{Anjos2020-kz}, and unfitted finite element methods \cite{Frachon2023-dn}; more comprehensive reviews can be found in \cite{Liu2018-ve, Soligo2019-jt, Mialhe2023-bp}. All of these models have explored the effect of surfactants in multicomponent flows where interfacial dynamics are solely driven by fluid motion. 

A much more challenging problem is understanding the effects of surfactants on interfaces where liquid-vapor phase transformations (also referred to as mass transfer) occur. Phase transformations are very important in the natural sciences and engineering. For example, due to its high specific heat capacity and latent heat of vaporization, liquid water is an excellent medium for transferring heat energy. The heat transfer rate is highly related to the interface dynamics. Studies have shown that the addition of surfactants enhances the heat transfer rate by reducing the interfacial energy required for bubble generation and by suppressing bubble coalescence. As a result, small nucleate bubbles are produced from the heated surface at a higher frequency, while the formation of large bubble clusters is diminished, leading to a more stable boiling process. Beyond the heat transfer process, erosion caused by pressure-induced cavitation poses a serious challenge for marine vehicles, hydro-turbines, gears in cooling liquids, and similar systems. In such situations, the liquid can hardly be considered pure water, as many impurities behave like surfactants and significantly influence the cavitation process. Simultaneously accounting for surfactant transport and phase change poses a significant modeling challenge.

Modeling liquid-vapor phase transformations computationally requires simultaneous consideration of non-equilibrium thermodynamics as well as large density and viscosity gradients. Most existing models are based on either a thermodynamic equilibrium assumption or source terms in the mass conservation equations that involve phenomenological parameters. The few studies \cite{Yiantsios2010-lx, Premnath2018-sz} that do couple surfactant transport with phase transformation adopt such source-term-based formulations. Yet, such parameter-dependent phase-changing models lack generalizability and may not be adequate for models that include surfactants unless they are recalibrated. 

This work proposes a phase-field model of surfactant-mediated liquid-vapor phase transformations. In the context of phase-transforming flows, employing nonconvex potentials has shown success in predicting interfacial dynamics and mass transfer in liquid-vapor systems. Liquid-vapor systems involve a non-convex bulk free energy and an interfacial free energy. The former has been used to predict thermodynamic properties and critical points  \cite{Korteweg1891-no}. The latter depends on the gradient of density, accounting for the effect of surface tension. The combination of the non-local free energy with the balance equations of compressible flows in a thermodynamically consistent manner, using the Coleman-Noll procedure, leads to the Navier-Stokes-Korteweg (NSK) equations \cite{Gomez2010-rz,Gomez2017-ip,Liu2013-kl}. The NSK equations comply with the second law of thermodynamics and have demonstrated their ability to model both pressure-induced and temperature-induced liquid-vapor phase transformations through Direct van der Waals simulations (DVS) \cite{Hu2023-ab,Hu2022-xx}.

Despite its success, the NSK framework has not been used to explore the effect of surfactants on liquid-vapor systems. In this work, we propose a thermodynamically consistent modification of the NSK equations that applies to soluble surfactants. We show that the proposed method successfully captures the variation of surface tension with respect to surfactant concentration. In addition, the method decouples surface tension from diffuse-interface thickness, thereby improving robustness across a wide range of simulation cases. Owing to its generality, the model offers substantial flexibility and can readily be coupled with other phase-transformation models and non-condensable gas effects \cite{Hu2023-ab,Mukherjee2019-cv,Mukherjee2020-si,Mukherjee2022-wd}.

The development of these methodologies will enhance our understanding of surfactant dynamics and expand their applications across various scientific and engineering disciplines.

\section{Method}

\subsection{Model overview}

We investigate the hydrodynamics of a fluid-surfactant mixture. The fluid undergoes liquid-vapor phase transformations. Among the various forms of liquid-vapor transformations, we are primarily interested in hydrodynamic cavitation, which can be considered an isothermal process in many cases of practical interest \cite{Mukherjee2025-bh}. Thus, we describe the phase-transforming fluid using the isothermal form of the NSK equations
\begin{alignat}{2}\label{eq:NSK}
        &\frac{\partial \rho}{\partial t} + \nabla\cdot(\rho \boldsymbol{u}) = 0, \quad \text{in }\Omega\\
        \label{eq:NSK-colm}
        &\frac{\partial (\rho \boldsymbol{u})}{\partial t} + \nabla\cdot(\rho \boldsymbol{u}\otimes \boldsymbol{u}) - \nabla\cdot \boldsymbol{T} = 0, \quad \text{in }\Omega
\end{alignat}
Here, $\Omega$ is the spatial domain, $\rho$ is the fluid density and $\boldsymbol{u}$ is the fluid velocity. The Cauchy stress tensor is $\boldsymbol{T} = -p\boldsymbol{I} + \boldsymbol{\tau} + \boldsymbol{\xi}$, where  $p$ is the fluid pressure, $\boldsymbol{\tau}$ is the viscous stress and $\boldsymbol{\xi}$ is the Korteweg stress. The viscous and Korteweg stress tensors are defined, respectively, as:
\begin{alignat}{2}
        &\boldsymbol{\tau} = \Bar{\mu}(\nabla \boldsymbol{u} + \nabla^{T}\boldsymbol{u}) - \frac{2}{3}\Bar{\mu} \nabla\cdot \boldsymbol{u} \boldsymbol{I},\\
        &\boldsymbol{\xi} = \lambda (\rho\nabla^2\rho + \frac{1}{2}|\nabla\rho|^2)I - \lambda \nabla\rho \otimes\nabla\rho,
\end{alignat}
where $\Bar{\mu}$ is the fluid's dynamic viscosity and $\lambda$ is a positive constant that is proportional to the square of the thickness of the liquid-vapor interface.

Surfactant molecules are amphiphilic, consisting of a polar hydrophilic head and a non-polar hydrophobic tail. In a liquid-vapor multiphase system, such as liquid water in contact with water vapor, the hydrophobic tails preferentially orient towards the vapor side of the interface, while the hydrophilic heads remain in the aqueous phase \cite{Rosen2004-qp}. Hence, we assume that the surfactants dissolve in the liquid bulk and adsorb at the interface, while their concentration in the vapor phase is negligible. To avoid solving an interfacial surfactant transport equation, we assume that adsorption and desorption are fast enough that the surfactant molecules rearrange between the interface and the bulk instantaneously. As a result, the surface concentration of surfactant $\Gamma$ varies rapidly with the local bulk concentration $\phi$, as described by the Langmuir adsorption isotherm $\Gamma(\phi)=\Gamma_\infty K \phi/(1+K\phi)$, where $\Gamma_\infty$ is the maximum surface concentration of surfactant and $K$ is the adsorption equilibrium constant. The Gibbs adsorption equation relates the surface tension $\sigma$ with $\phi$ through $\dd \sigma=-R_g\theta\Gamma(\phi)\dd\,\!\ln(\phi)$, where $R_g$ is the gas constant and $\theta$ is the temperature. Combining these two relations, we have the closed expression for surface tension directly as a function of bulk concentration, which is $\sigma(\phi)=\sigma_0 - R_g \theta \Gamma_\infty\ln(1+K\phi)$. Here $\sigma_0$ denotes the surface tension of pure liquid in the absence of surfactant.

Because the surfactant concentration is negligible in the vapor phase, we solve surfactant transport equation on ${\Omega}_l(t)$, i.e., the time-dependent spatial domain, which excludes vapor. The surfactant transport equation is
\begin{equation}\label{eq:moving_domain}
    \frac{\partial \phi }{\partial t} + \nabla\cdot (\phi \boldsymbol{u}) - \nabla \cdot (D_f\nabla \phi)=0, \quad \text{in }{\Omega}_l(t),
\end{equation}
where $D_f$ is the molar diffusivity of the surfactant. Solving Eq.~\eqref{eq:moving_domain} is challenging as the computational domain changes over time. To circumvent this issue, we use the diffuse-domain approach \cite{Bures2021-tc}. The diffuse-domain approach allows us to define an equation that is equivalent to~\eqref{eq:moving_domain}, but is defined on the domain $\Omega$, which is fixed in time. Note that $\Omega$ is the domain where the NSK equations are defined [Eq.~\eqref{eq:NSK}], and we have $\Omega\supset\Omega_l(t)$ for all $t$. To redefine Eq.~\eqref{eq:moving_domain} in $\Omega$, we introduce a dimensionless function $g(\rho)$ that localizes the liquid phase and is defined as
\begin{equation}\label{eq:g-term}
    g(\rho)=
    \begin{cases}
        2\epsilon, & \rho< \rho_v, \\
        g_\text{mid}(\rho), & \rho_v \le\rho\le\rho_l,\\
        1, & \rho>\rho_l,
    \end{cases}
\end{equation}
where
\begin{equation}
    g_\text{mid}(\rho) = 0.5+\epsilon+(0.5-\epsilon)\sin{\left[\frac{\rho-\rho_v}{\rho_l-\rho_v}\pi - \frac{\pi}{2}\right]}.
\end{equation}
Here, $\rho_l$ and $\rho_v$ represent the equilibrium liquid and vapor phase densities of the phase-transforming fluid, while $\epsilon$ is a small, dimensionless positive number ($5\times10^{-9}$ in our calculations) that is used to ensure numerical stability and avoid trivial solutions. 
The surfactant dynamics equation in $\Omega$ is
\begin{equation}\label{eq:COMC}
    \frac{\partial c }{\partial t} + \nabla\cdot (c \boldsymbol{u}) - \nabla \cdot \left[D_f(\nabla c-\phi\nabla g(\rho))\right]=0, \quad \text{in }\Omega,
\end{equation}
where $c=g(\rho)\phi$. In what follows, we refer to $c$ as the surfactant concentration for simplicity, and we use $c_l$ to represent the value of $c$ in the bulk liquid phase. We start the simulations with a uniform concentration in the bulk liquid, so that $\phi=c_l$ in the liquid phase.

\subsection{Thermodynamics and surface tension}
The total Helmholtz free energy functional for the system is
\begin{equation}\label{eq:Helmholtz}
A[\rho]=\int_\Omega\left[\Psi^\text{EoS}(\rho)+\frac{\lambda}{2}|\nabla\rho|^2\right]\mathrm{d}\boldsymbol{x},
\end{equation}
where $\Psi^\text{EoS}(\rho)$ is the bulk Helmholtz free energy density from the equation of state (EoS), the gradient term represents the free energy at the liquid-vapor interface, and $\boldsymbol{x}=(x,y)$ represents a spatial point in $\Omega$. We use the van der Waals (vdW) model for the bulk Helmholtz free energy density. For vdW fluids under isothermal conditions,  $\Psi^{\rm EoS}(\rho) = R_g\theta\rho\log(\rho/(b-\rho))-a\rho^2 $, and the chemical potential and pressure are defined as
\begin{alignat}{2}
\mu^\text{EoS}(\rho) &=  \frac{\dd\Psi^\text{EoS}(\rho)}{\dd\rho} \nonumber\\
                    &= R_g\theta\log\left(\frac{\rho}{b-\rho}\right)+R_g\theta\frac{b}{b-\rho}-2a\rho \label{eq:mu} \\
p^\text{EoS}(\rho) &= \rho\mu^\text{EoS}(\rho)- \Psi^\text{EoS}(\rho)\nonumber\\
&=R_gb\theta\frac{\rho}{b-\rho}-a\rho^2 \label{eq:eos},
\end{alignat}
where $a$ and $b$ are constants that define the fluid. 

We call the saturation densities of the liquid and vapor phases $\rho_l$ and $\rho_v$, respectively. The saturation densities are defined by the coexistence of the liquid and vapor phases at phase equilibrium ($\mu^\text{EoS}(\rho_v) = \mu^\text{EoS}(\rho_l)\equiv\mu_\text{sat}$) and mechanical equilibrium ($p^\text{EoS}(\rho_v) = p^\text{EoS}(\rho_l)\equiv p_\text{sat}$). For an isothermal system at saturation conditions, the total Helmholtz free energy of the system is minimized, subject to the conservation of mass, leading to the coexistence of the two phases. For a two-dimensional system with a planar interface parallel to $x$, the equilibrium density profile along the $y$ direction minimizes $A[\rho]$ subject to mass conservation. The Euler-Lagrange equation is
\begin{equation}\label{eq:EL-eqn}
    0=\mu^\text{EoS}(\rho)-\lambda\frac{\dd^2\rho}{\dd y^2}-\mu_\text{sat}
\end{equation}
where we have assumed that $\lambda$ is a constant. Now, we multiply Eq.~\eqref{eq:EL-eqn} by $\mathrm{d}\rho/\mathrm{d}y$ and integrate between $0$ and $y$. At $y=0$, we use the boundary conditions $\rho=\rho_v$, $\mathrm{d}\rho/\mathrm{d}y=0$. Thus,
\begin{equation}\label{eq:excessPsi}
        \frac{\lambda}{2}\left(\frac{\dd\rho}{\dd y}\right)^2 = 
        \Psi^\text{EoS}(\rho)- \Psi_\text{coex}(\rho),
\end{equation}
where 
\begin{equation}
\Psi_\text{coex}(\rho)=\Psi^\text{EoS}(\rho_v)+\mu_\text{sat}(\rho-\rho_v)
\end{equation}
is the linear function of $\rho$ that defines the common tangent to the saturation points in the $\rho$-$\Psi^\text{EoS}$ plane. The excess Helmholtz free energy $\Delta\Psi(\rho)$ is defined as 
\begin{equation}
\label{eq:excessPsi2}
\Delta\Psi(\rho)=\Psi^\text{EoS}(\rho)- \Psi_\text{coex}(\rho).
\end{equation}
In the interfacial region, $\Delta\Psi(\rho)>0$, because a homogeneous liquid vapor mixture with density $\rho\in(\rho_v,\rho_l)$ is thermodynamically unstable or metastable. 

Surface tension, $\sigma$, is the excess Helmholtz free energy per unit interfacial area required to maintain the diffuse interface. Mathematically:
\begin{equation}\label{eq:sigma_1}
    \sigma=\frac{1}{s}({A[\rho]}-{A}_\text{coex}[\rho]),
\end{equation}
where $s$ is the interface area and ${A}_\text{coex}[\rho]=\int_\Omega\Psi_\text{coex}(\rho)\mathrm{d}\boldsymbol{x}$ is the total Helmholtz free energy of a sharp-interface system. For a 1D planar interface lying on the $x$ plane, we have $\rho(\boldsymbol{x})=\rho(y)$, $\nabla\rho=(0,\frac{\dd\rho}{\dd y})$, and $\int_\Omega(\cdot)\dd\boldsymbol{x}=s\int_{-\infty}^{\infty}(\cdot)\dd y$, so Eq.~\eqref{eq:sigma_1} becomes 
\begin{equation}\label{eq:surfacetension-definition}
\sigma = \int_{-\infty}^{\infty} \Bigl[
        {\Psi^\text{EoS}(\rho)
                    + \frac{\lambda}{2}\Bigl(\frac{\dd\rho}{\dd y}\Bigr)^2}
        -{\Psi_{\text{coex}}(\rho)}
        \Bigr]\,
        \dd y.
\end{equation}
Using Eqs.~\eqref{eq:excessPsi} and \eqref{eq:excessPsi2}, we can define the change of integration variable $\dd y = \sqrt{\frac{\lambda}{2}}\frac{\dd\rho}{\sqrt{\Delta\Psi(\rho)}}$. Because a 1D planar interface satisfies $\rho(y\rightarrow\infty)=\rho_l$ and $\rho(y\rightarrow-\infty)=\rho_v$, we obtain:
\begin{equation}\label{eq:surfacetension-equation}
    \begin{split}
        \sigma &=\int_{\rho_v}^{\rho_l}\sqrt{2\lambda\Delta\Psi(\rho)}\dd\rho\\ 
        &= \int_{\rho_v}^{\rho_l}\sqrt{2\lambda[\Psi^\text{EoS}(\rho)-\Psi^\text{EoS}(\rho_v)-\mu_{\text{sat}}(\rho-\rho_v)]}\dd\rho.
    \end{split}
\end{equation}

\subsection{Surfactant modulation of surface tension}

Fluid molecules at a fluid-fluid interface experience an imbalance in cohesive forces due to the asymmetry in molecular properties across the interface, resulting in higher energy compared to the bulk phase. In water, strong hydrogen bonding produces a high interfacial energy between the liquid and vapor phases, which results in surface tension. Surfactant molecules accumulate at the liquid-vapor interface and disrupt cohesive interactions. Hence, they reduce the excess Helmholtz free energy density required to sustain non-equilibrium densities in the interfacial region. To model this in a thermodynamically-consistent manner, we propose a reconstruction of $\mu^\text{EoS}(\rho)$ that decreases $\Delta\Psi(\rho)$ in the interfacial region. The strength of the reduction of $\Delta\Psi(\rho)$ depends on the local surfactant concentration $c$. The reduction of excess free energy is achieved by replacing $\mu^\text{EoS}(\rho)$ with $\mu^\text{rc}(\rho,c)$. All other thermodynamic quantities are changed accordingly using Eq.~\eqref{eq:mu} and Eq.~\eqref{eq:eos}. We call $\Psi^\text{rc}(\rho,c)$ and $p^\text{rc}(\rho, c)$ the modified (reconstructed) free energy and pressure, respectively. From the first identity in Eq.~\eqref{eq:surfacetension-equation}, we can see that the reconstruction will lead to a reduction in the surface tension $\sigma$. 

\begin{figure}
    \centering
    \includegraphics[width=\columnwidth]{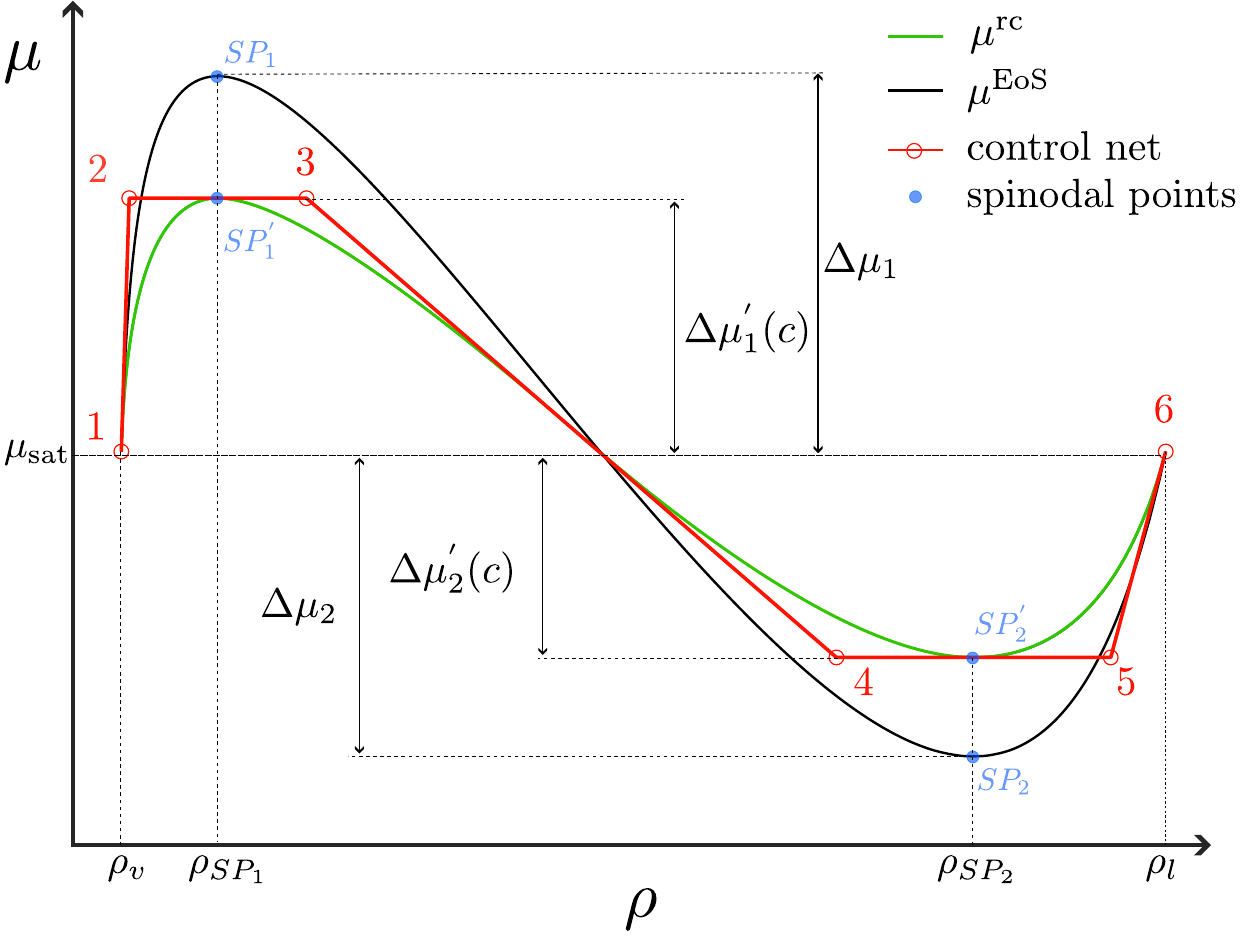}
    \caption{Original (black) and reconstructed (green) chemical potential in the interfacial region.}
    \label{fig:1st}
\end{figure}

We define the reconstructed chemical potential $\mu^\text{rc}(\rho,c)$ using the method proposed in \cite{Mukherjee2023-ke}. We use a B-spline function with carefully chosen control points to preserve the bulk-equilibrium characteristics and inherent non-convexity of the vdW EoS, while ensuring global differentiability for numerical stability. It provides a smooth reconstruction of the non-monotone chemical-potential curve with only a few parameters, enabling incorporation of the surfactant effect on surface tension.
Fig.~\ref{fig:1st} compares the original and modified chemical potentials. At a fixed subcritical temperature, the vdW chemical potential $\mu^\text{EoS}(\rho)$ is nonconvex and exhibits two spinodal points where $\dd\mu^\text{EoS}(\rho)/\dd\rho=0$. These points are denoted by $SP_1$ and $SP_2$ in Fig.~\ref{fig:1st} and their $(\rho,\mu)$ coordinates are $(\rho_{SP_1}, \mu_{\text{sat}}+\Delta\mu_1)$ and $(\rho_{SP_2}, \mu_{\text{sat}}-\Delta\mu_2)$, respectively. The inequality $\Delta\mu_1\neq\Delta\mu_2$ reflects the asymmetry of the vdW EoS.

We use a second-order B-spline with 6 control points to construct $\mu^{\text{rc}}(\rho,c)$ for $\rho\in[\rho_v, \rho_l]$. Points 1 and 6 are located at equilibrium points to ensure the global continuity of the chemical potential. The $\mu$-values  of points 2 and 3 are defined by 
\begin{equation}
    \Delta\mu_1'(c)=\Delta\mu_1/\eta(c), \quad \eta(c)\geq1
\end{equation}
The $\mu$-values for points 4 and 5 are defined by 
\begin{equation}
    \Delta\mu_2'(c)=\kappa(c)\Delta\mu_1'(c).
\end{equation}
For each value of $c$, the reduction factors $\eta(c)$ and $\kappa(c)$ are determined iteratively so that two conditions are verified: 1) the difference between the model-predicted surface tension and the corresponding experimentally measured surface tension is minimum; and 2) the Maxwell equal area rule is satisfied, i.e.,  $\int_{\rho_v}^{\rho_l}(\mu^{\text{rc}}(\rho,c)-\mu_\text{sat})\dd\rho$=0. For each prescribed concentration $c_0$, the value of $\eta(c_0)$ is adjusted to reduce the surface tension; for each $\eta(c_0)$, there is a unique $\kappa(c_0)$ that satisfies the Maxwell equal area rule. The pair $(\eta(c_0), \kappa(c_0))$ is accepted once both conditions are met. After repeating this operation for multiple values of $c$, we fit polynomials to the data to ensure differentiability of $\eta$ and $\kappa$ with respect to $c$. 

The $\rho$-coordinates of points 2 and 5 are selected to ensure that $\mu^{\text{rc}}(\rho,c)$ is differentiable at the equilibrium points. The $\rho$-coordinates of points 3 and 4 are chosen so that the spinodal points $SP_1'$ and $SP_2'$ in $\mu^{\text{rc}}(\rho,c)$ remain at $\rho_{SP_1}$ and $\rho_{SP_2}$, respectively. 

After the reconstruction of $\mu^{\text{rc}}(\rho,c)$, we can define the reconstructed free energy $\Psi^{\text{rc}}(\rho, c)$ and pressure $p^\text{rc}(\rho, c)$ using standard thermodynamic relations:
\begin{align}\label{eq:reconstructed_phi_p}
    &\Psi^{\text{rc}}(\rho, c) = \Psi^{\text{EoS}}(\rho_v) + \int_{\rho_v}^{\rho} \mu^{\text{rc}}(\rho^*, c)\mathrm{d}\rho^*,\\
    &p^\text{rc}(\rho,c) = \rho \mu^{\text{rc}}(\rho, c) - \Psi^{\text{rc}}(\rho,c).
\end{align}
%
The pressure field that enters the NSK equations (Eq.~\eqref{eq:NSK-colm}) is
\begin{align}
p(\rho, c) = \begin{cases}
    p^{\text{rc}}(\rho,c), & \text{if}\; \rho_v<\rho<\rho_l\\
    p^{\text{EoS}}(\rho), & \text{otherwise}
\end{cases}
\end{align}
Because the reduction of chemical potential will lead to a thicker interface \cite{Jamet2001-wr}, the parameter $\lambda$, which controls the interface thickness, needs to be re-scaled accordingly to vary surface tension without altering the interface thickness. We change the interfacial parameter using  the constant bulk concentration $c_l$,
\begin{equation}
    \lambda=\lambda_0/\eta(c_l),
\end{equation}
where $\lambda_0$ is the positive constant that is proportional to the square of the interface thickness for clean water. After the reconstruction, the surface tension is related to the concentration of surfactant through the equation
\begin{equation}
    \sigma = \int_{\rho_v}^{\rho_l}\sqrt{2\lambda[\Psi^{\text{rc}}(\rho, c)-\Psi^{\text{EoS}}(\rho_v)-\mu_{\text{sat}}(\rho-\rho_v)]}\dd\rho \\
\end{equation}
This rescaling, in addition to being robust and applicable to a wide range of surfactants,  ensures that the interface thickness is independent of $c$.
\subsection{Dimensionless form of the governing equations}
Prior to discretization, we rewrite the governing equations in non-dimensional form. We follow \cite{Gomez2010-rz} and scale length by $L_0$, mass by $bL_0^3$, time by $L_0/\sqrt{ab}$ and temperature by $\theta_c=8ab/(27R_g)$, which is the critical temperature of a vdW fluid. Surfactant concentration is scaled by $b/M$, where $M$ is the molar mass of the phase-transforming fluid. The non-dimensional form of the governing equations is
\begin{align}
\label{eq:continuity_dimensionless}
    &\frac{\partial \hat{\rho}}{\partial \hat{t}} + \hat{\nabla}\cdot(\hat{\rho} \hat{\boldsymbol{u}}) = 0 \\
\label{eq:momentum_dimensionless}
    &\frac{\partial (\hat{\rho} \hat{\boldsymbol{u}})}{\partial \hat{t}} + \hat{\nabla}\cdot(\hat{\rho} \hat{\boldsymbol{u}}\otimes \hat{\boldsymbol{u}} + \hat{p} \boldsymbol{I}) - \frac{1}{\text{Re}}\hat{\nabla}\cdot \hat{\boldsymbol{\tau}} - \frac{\text{Ca}^2}{\eta(c_l)}\hat{\rho}\hat{\nabla}(\hat{\nabla^2}\hat{\rho}) = 0\\
\label{eq:surfactant_dimensionless}
    &\frac{\partial \hat{c} }{\partial \hat{t}} + \hat{\nabla}\cdot (\hat{c} \hat{\boldsymbol{u}}) - \frac{1}{\text{Pe}}\hat{\nabla} \cdot \left[(\hat{\nabla} \hat{c}-\frac{\hat{c}}{g(\hat{\rho})}\hat{\nabla} g(\hat{\rho}))\right]=0\\
\label{eq:pressure_dimensionless}
    &\hat{p} = \frac{8}{27}\frac{\hat{\theta}\hat{\rho}}{1-\hat{\rho}}-\hat{\rho}^2\\
\label{eq:viscosity_dimensionless}
&\hat{\boldsymbol{\tau}}=\hat{\nabla}\hat{\boldsymbol{u}}+\hat{\nabla}^{T}\hat{\boldsymbol{u}}-\frac{2}{3}\hat{\nabla}\cdot\hat{\boldsymbol{u}}\,\boldsymbol{I},
\end{align}
where we used a hat over symbols to represent non-dimensional quantities. The Reynolds number, capillary number, and Peclet number are defined as:
\begin{equation}\label{eq:ReCaPe}
    \text{Re}=\frac{L_0 b \sqrt{ab}}{\bar{\mu}} , \quad
    \text{Ca}=\frac{\sqrt{\lambda_0/a}}{L_0}
    , \quad
    \text{Pe} = \frac{L_0\sqrt{ab}}{D_f}.
\end{equation}
In what follows, we will report dimensionless results. Compatible data for dimensional quantities could be obtained using the following values for water: The temperature is $\theta =300 \si{K}$, the molar mass of water is $M=18\si{g/mol}$, and the gas constant is $R_g=\frac{8314.46}{18}\si{J/(kg\cdot K)}$. The vdW EoS constants are: $a=1.7088\times10^{3} \si{m^5/(kg\cdot s^2)}$, $b=590 \si{kg /m^3}$, and the critical temperature $\theta_c=\frac{8}{27}\frac{ab}{R_g}=647.2\si{K}$. In the subsequent text, we present all equations in dimensionless form. For simplicity, we will drop the hat notation. Any quantity written without units is understood to be dimensionless. Dimensional values are used only when their units are provided explicitly, such as $\si{mM}\equiv\si{mol/m^3}$ for surfactant concentration and $\si{mN/m}$ for surface tension. To facilitate comparison with experimental data, we report $c$ and $\sigma$ in dimensional form with units when appropriate.

\subsection{Numerical discretization}

In preparation for an isogeometric discretization \cite{Hughes2005-hc}, we derive a weak form of the problem. For simplicity, we assume periodic boundary conditions, but all other boundary conditions used in the paper can be imposed in a straightforward manner using standard procedures \cite{Hughes1987-cm}. Let $\boldsymbol{X}$ denote the space that serves both as the solution space and as the weighting function space. The variational problem is: Find $\boldsymbol{U} = \{\rho, \boldsymbol{u}, c\} \in \boldsymbol{X}$ such that for all $\boldsymbol{W} = \{w_1, \boldsymbol{w}_2, w_3\} \in \boldsymbol{X}$, $B(\boldsymbol{W}, \boldsymbol{U}) = 0$, where
\begin{alignat}{2}
    B & = \left(w_1, \frac{\partial\rho}{\partial t}\right)_{\Omega} + \left( \boldsymbol{w}_2, \rho\frac{\partial \boldsymbol{u}}{\partial t}\right)_{\Omega}+ \left(\boldsymbol{w}_2, \boldsymbol{u}\frac{\partial \rho}{\partial t}\right)_{\Omega} \nonumber\\ 
    &+ \left( w_3, \frac{\partial c}{\partial t} \right)_{\Omega}-\left( \nabla w_1, \rho\boldsymbol{u}\right)_{\Omega}-\left( \nabla\boldsymbol{w}_2,\rho \boldsymbol{u}\otimes\boldsymbol{u}\right)_{\Omega} \nonumber\\ 
    &- \left(\nabla\cdot\boldsymbol{w}_2, p\right)_{\Omega}+\frac{1}{\text{Re}}\left(\nabla \boldsymbol{w}_2, \nabla \boldsymbol{u} +\nabla\boldsymbol{u}^{T}\right)_{\Omega}\nonumber\\
    &-\frac{2}{3}\frac{1}{\text{Re}}\left(\nabla\cdot\boldsymbol{w}_2, \nabla \cdot\boldsymbol{u}\right)_{\Omega}+\frac{\text{Ca}^2}{\eta(c_l)}\left(\nabla\cdot\boldsymbol{w}_2, \rho\nabla^2\rho\right)_{\Omega}\nonumber\\ 
    &+\frac{1}{2\eta(c_l)}\text{Ca}^2\left(\nabla\cdot\boldsymbol{w}_2, |\nabla\rho|^2\right)_{\Omega}-\frac{\text{Ca}^2}{\eta(c_l)}(\nabla\boldsymbol{w}_2,\nabla\rho\otimes\nabla\rho)_{\Omega} \nonumber\\ 
    &- \left( \nabla w_3, c\boldsymbol{u}\right)_{\Omega}+\frac{1}{\text{Pe}}\left(\nabla w_3,\nabla c - \frac{c}{g(\rho)}\nabla g(\rho) \right)_{\Omega} 
    \label{weakform}
\end{alignat}
The discretization of the weak form in Eq.~\eqref{weakform} requires globally smooth basis functions, which can be obtained using Isogeometric Analysis for the spatial discretization of the problem. In this study, we use Nonuniform Rational B-Splines (NURBS). We integrate in time using the generalized-$\alpha$ method. 

\section{Results}

\subsection{Surface tension predictions based on Laplace pressure}\label{LP}

\begin{figure}
    \centering
    \includegraphics[width=\columnwidth]{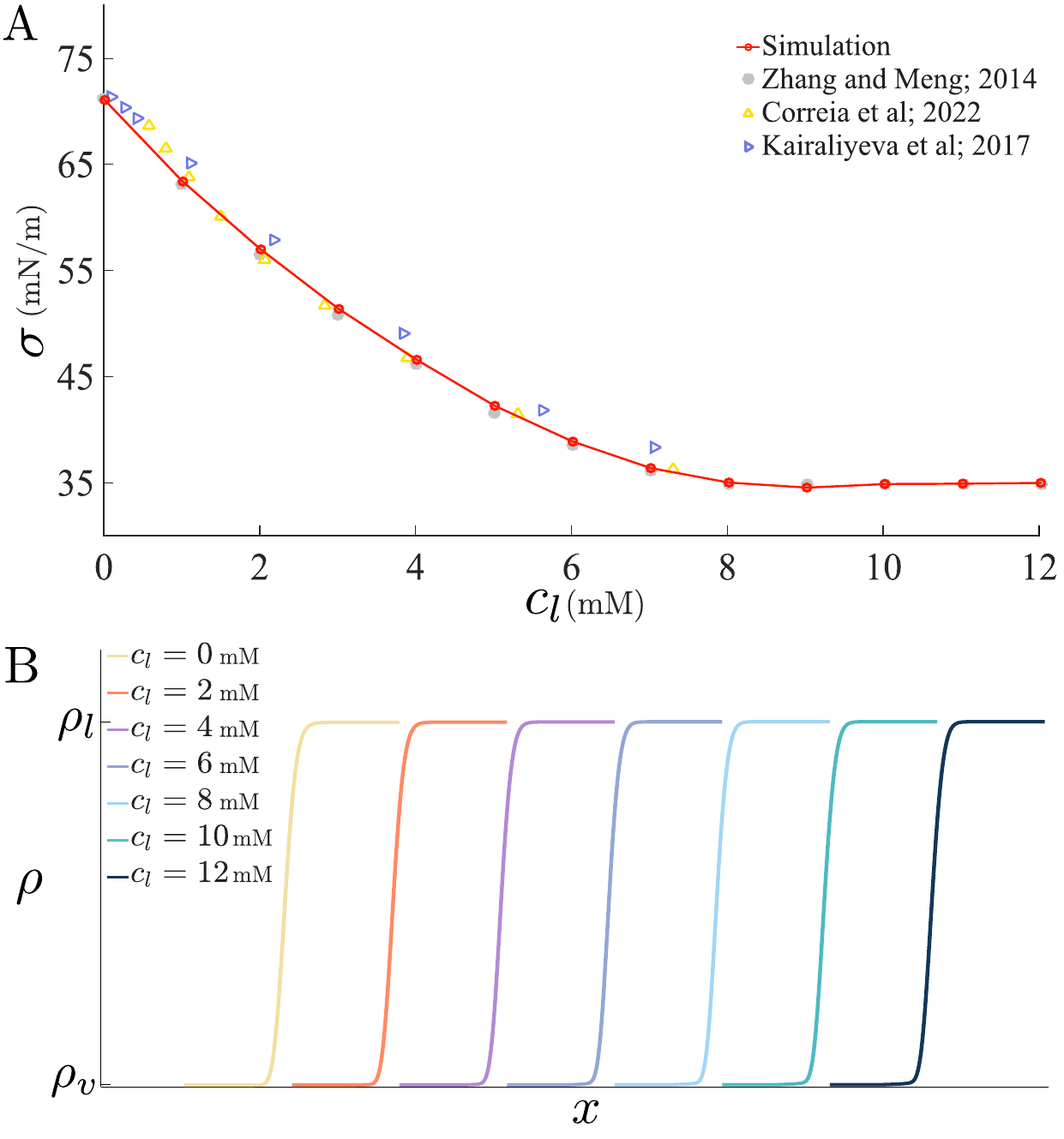}
    \caption{(A) Surface tension obtained from simulation (red line) and experiments \cite{Zhang2014-zp, Correia2022-ys, Kairaliyeva2017-ez} as a function of SDS concentration. (B) Spatial variation of the density for different values of the surfactant concentration, showing that the density profile does not change with surfactant concentration.}
    \label{fig:laplace_pressure_tot}
\end{figure}

While our model is applicable to any soluble surfactant, this section will focus on Sodium Dodecyl Sulfate (SDS) to allow comparisons with experiments. SDS is highly soluble in water, widely used, and well characterized \cite{Zhang2014-zp,Al-Soufi2012-ct,Correia2022-ys, Kairaliyeva2017-ez}. The relation between surface tension and surfactant concentration in aqueous SDS solutions is shown in Fig.~\ref{fig:laplace_pressure_tot}(A). The calibration is based on one experimental study \cite{Zhang2014-zp}, and comparisons with other sources are also provided. The critical micelle concentration (CMC) is about 8.1 mM, and the surface tension does not decrease further with increasing concentrations above the CMC.  

To demonstrate the model's capability of accurately capturing the variation of surface tension with SDS concentration, we perform a series of numerical simulations to obtain the surface tension values at different concentration levels. In this section, we will simulate cylindrical (2D) bubbles under equilibrium and estimate surface tension from the Young-Laplace equation
\begin{equation}\label{eq:YLaplace}
    \sigma = R\Delta p= R(p_v - p_l).
\end{equation}
Here, $\Delta p$ is the Laplace pressure, $p_v$ the pressure in the vapor phase and $p_l$ the pressure in the liquid phase for a cylindrical bubble in equilibrium. To evaluate the surface tension obtained in the simulations, we will use the simulation results to get values of $p_v$, $p_l$ and $R$. The pressures $p_v$ and $p_l$ can be directly obtained from the simulation data. To obtain the value of $R$, we plot
the density profile along a line that crosses the bubble center and compute
\begin{equation}
    R = \int_0^{\infty} r\left(\frac{\mathrm{d}\rho}{\mathrm{d}r}\right)^2\mathrm{d}r\Big/\int_0^{\infty}\left(\frac{\mathrm{d}\rho}{\mathrm{d}r}\right)^2\mathrm{d}r.
\end{equation}
where $r=|\boldsymbol{x}-\boldsymbol{x}_c|$ is the radial distance from the bubble center. 

The simulations start with a vapor bubble of radius $R=0.15$ and density $\rho_v$, centered at $\boldsymbol{x}=(0.5,0.5)$ in a square computational domain whose spatial coordinates satisfy $x,y\in[0, 1]$. The surrounding liquid has a density of $\rho_l$. The initial density field is prescribed using the hyperbolic tangent function in Cartesian coordinates, 
\begin{equation}
    \rho(x,y)=\frac{\rho_v+\rho_l}{2} + \frac{\rho_l-\rho_v}{2}\tanh{\frac{\mathcal{X}_d-R}{\text{Ca}}},
\end{equation} 
where $\mathcal{X}_d=\sqrt{(x-0.5)^2 + (y-0.5)^2}$. In each simulation, the value $\phi=c_l$ is set uniformly throughout the entire domain, and using the localization field $g(\rho)$ in Eq.~\eqref{eq:g-term}, the initial condition of concentration is defined as $c=g(\rho)\phi$. Periodic boundary conditions are imposed in both directions. 

To establish parameter values related to flow that can be resolved on a given mesh, we follow the scaling proposed in \cite{Gomez2010-rz}
\begin{equation}\label{eq:ReCaPe_new}
   \text{Ca}=\frac{h}{L_0}, \quad \text{Re}=2\text{Ca}^{-1},\quad \text{Pe}= \frac{\sqrt{\lambda_0 b}}{\text{Ca}D_f}
\end{equation}
where $h=L_0/N$ is the characteristic length scale of a uniform spatial mesh, and $N$ is the number of elements along each direction. This scaling ensures the interface thickness remains well-resolved by the mesh and that the Weber number stays of order unity, i.e., $\text{We}=\text{Ca}\cdot \text{Re}\sim O(1)$. Here, we choose $N=256$, such that $\text{Ca}=1/256$, $\text{Re}=512$. We took $\text{Pe}=145$. 

The initial density distribution does not correspond to a perfect equilibrium state. Thus, when the simulation starts, the bubble shrinks and grows periodically until the system reaches equilibrium. To quantitatively monitor how far the simulation is from equilibrium, we evaluate the fluid speed over the domain as
\begin{equation}
    U (t) = \int_\Omega |\boldsymbol{u}(\boldsymbol{x},t)|\dd \boldsymbol{x},
\end{equation}
and track the relative change $\zeta=(U(t_{i+1})-U(t_{i}))/U(t_{i})$, where $\Delta t = t_{i+1}-t_i$ is the time step. We consider the system to have reached equilibrium when $\zeta\leq10^{-5}$. 


Fig.~\ref{fig:laplace_pressure_tot}(A) compares the surface tension values obtained from simulations with those obtained from experiments \cite{Zhang2014-zp}. The model captures the decreasing trend with increasing surfactant concentration $c$ up to the CMC and the subsequent plateau. 
\begin{table}
    \centering
    \begin{tabular}{c|c|c}
        \toprule
         $c_l$ ($\si{mM}$) &  $n$ & $\delta\times10^2$ \\
         \midrule
         0 & 688 & 1.601 \\
         2 & 688 & 1.598 \\
         4 & 684 & 1.594 \\
         6 & 684 & 1.581 \\
         8 & 684 & 1.573 \\
         10 & 684 & 1.565 \\
         12 & 684 & 1.560 \\
         \bottomrule
    \end{tabular}
    \caption{Liquid-vapor interface thickness ($\delta$) with different surfactant concentration levels, and the number of elements ($n$) used to discretize the interface.}
    \label{tab:interface_thickness}
\end{table}

One advantage of the proposed model over the model in \cite{Bueno2016-al} is that in the model proposed herein the interface thickness is independent of different surfactant concentrations. This can be seen in Fig.~\ref{fig:laplace_pressure_tot}(B), which shows the density field across the vapor bubble at equilibrium for different values of $c_l$. The interface thickness looks the same for all values of $c_l$. To make the comparison more quantitative, we proceed as follows: we count the number of elements that fall in the interfacial region ($n$) and compute an estimate of the interface thickness ($\delta$). All computations were performed on the same uniform mesh, so $n$ is an estimate of the interfacial area. To determine $n$, we run the simulations up to steady state and then define the bulk density plateaus $\bar{\rho_v}=\text{pct}_2(\rho)$ and $\bar{\rho_l}=\text{pct}_{98}(\rho)$, where $\text{pct}_2$ and $\text{pct}_{98}$ represent the 2\textsuperscript{nd} and 98\textsuperscript{th} percentiles of the element-wise density distributions, respectively. Then, we define the density range $\Delta\bar{\rho}=\bar{\rho_l}-\bar{\rho_v}$. We label the elements whose center density is in the range $[\bar{\rho_v}+0.05\Delta\bar{\rho}, \bar{\rho_l}-0.05\Delta\bar{\rho}]$ as interfacial cells and count the total number of them ($n$). The interfacial thickness is defined as $\delta=n h^2 / (2\pi R)$. The results are shown in Table~\ref{tab:interface_thickness}. The interface thickness is fairly constant across a large range of surfactant concentrations. By controlling the interface thickness, we sidestep the challenges associated with the uncertain physical implications of changes in interface thickness. In addition, we can a priori determine the minimal element size required to perform a stable simulation.

\subsection{Surface tension predictions based on oscillation frequency of a liquid-vapor interface}

\begin{figure}
    \centering
    \includegraphics[width=\linewidth]{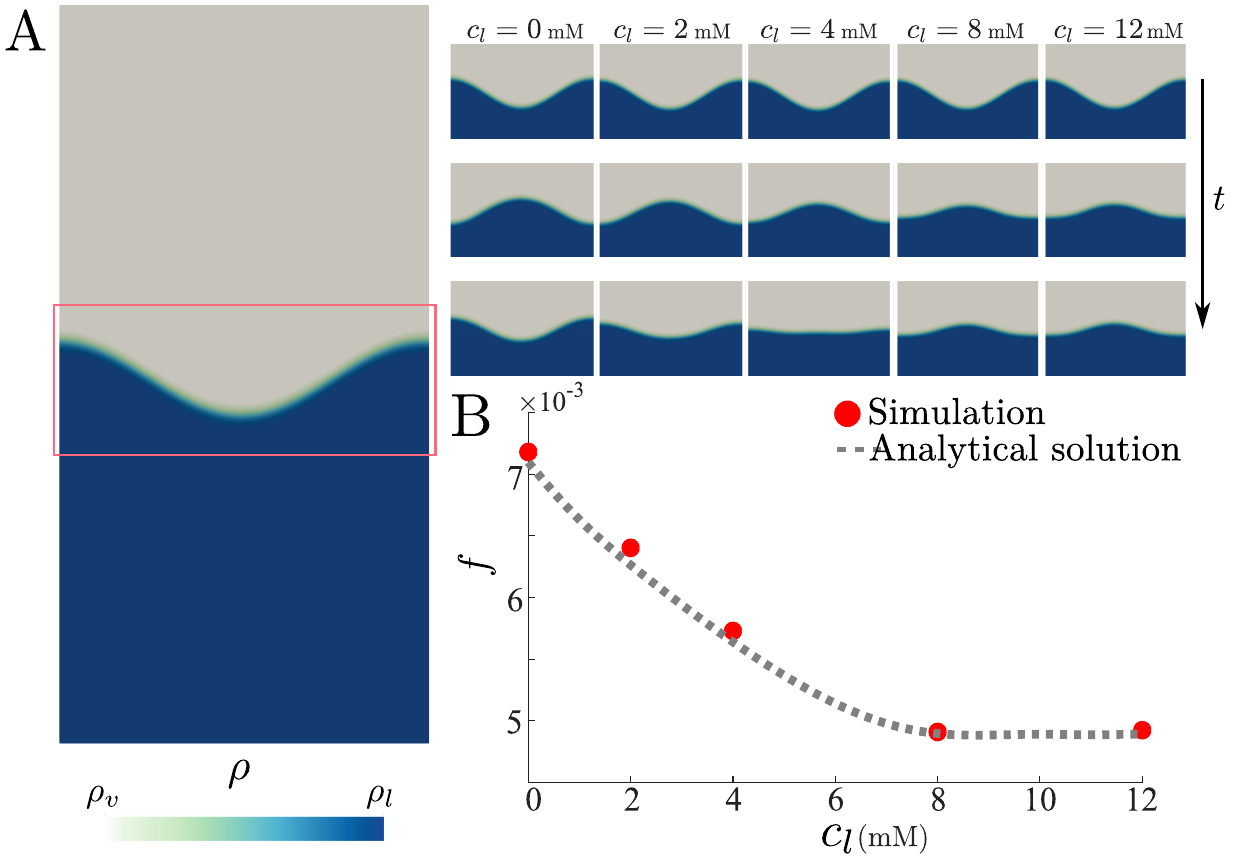}
    \caption{(A) Computational domain showing liquid water at the bottom and water vapor at the top. The initial interface has a sinusoidal shape. The red box highlights the region of interest, whose time evolution is shown on the right for different values of $c_l$. (B) Oscillation frequency of the interface as a function of surfactant concentration.}
    \label{fig:interface_oscillation}
\end{figure}

The results in Sect.~\ref{LP} show that the model can capture surfactant-induced surface tension variations in an equilibrium situation. We now assess its performance under non-equilibrium conditions. We accomplish this by studying the oscillation of a liquid-vapor interface. The oscillation frequency depends on surface tension and serves as an indirect measure of it. Fig.~\ref{fig:interface_oscillation}A illustrates the system that we use to study the liquid-vapor interface oscillation. We use a rectangular computational domain $\Omega=(0,0.5)\times(0,1)$ discretized into $256\times512$ elements. The parameters used in these simulations are $\text{Re}=5000, \text{Ca}=1/64, \text{Pe}=36$. We assume periodic boundary conditions on the left and right sides. On the top and bottom sides, we set no-slip boundary conditions for the velocity. At $t=0$, the fluid velocity is zero. The liquid-vapor interface is curved and is defined by the following initial condition for the density field
\begin{equation}
    \rho(y)=\frac{\rho_l+\rho_v}{2}+\frac{\rho_l-\rho_v}{2}\tanh{\frac{\mathcal{X}-y}{\text{Ca}}},
\end{equation}
where $\mathcal{X} = 0.5+0.05\cos{(kx)}$, and $k=4\pi$ is the wave number of the capillary wave. The initial condition for the surfactant concentration is $c=gc_l$. From this initial condition, the system evolves with a periodic interface oscillation driven by inertia and surface tension. Under the assumptions listed in \cite{Lautrup2004-xb} the oscillation frequency $f$ can be theoretically estimated by 
\begin{equation}\label{eq:analytical_frequency}
    f=\frac{1}{2\pi}\sqrt{\frac{\sigma}{\rho}k^3},
\end{equation}
A larger $\sigma$ exerts a stronger restoring force to diminish the curvature more rapidly. 


In Fig.~\ref{fig:interface_oscillation}B, we plot the frequency from Eq.~\eqref{eq:analytical_frequency} as a function of $c_l$ using a gray dashed line. The relation between $\sigma$ and $c_l$ was obtained from the experiments in \cite{Zhang2014-zp}. The simulation data is plotted with red circles. To obtain the values of the frequency from the simulation data, we monitor the temporal variation of density at the midpoint, i.e., $(x,y)=(0.5,1)$, log every instance when the interface ($\rho_\text{mid}=(\rho_v+\rho_l)/2$) crosses the midpoint, and compute the oscillation period as two times the time difference between two consecutive crossings. 

The oscillation frequency decreases as the surfactant concentration increases, indicating a slower stabilization process due to the diminished restoring force from lower surface tension. The good agreement between the theoretical solution and the simulation results confirms the ability of the proposed model to capture surface tension variations in response to surfactant concentration under nonequilibrium conditions.

\subsection{Effect of surfactant on bubble deformation under shear flow}
In the absence of flow and body forces, bubbles and droplets maintain a spherical shape that minimizes their surface area for a given volume. However, when subjected to flow, they deform and lose their spherical shapes. Lower surface tension makes bubbles and droplets more prone to deformation. Here, we use our model to analyze the effect of surfactant on bubble deformation under shear flow.

\begin{figure}
    \centering
    \includegraphics[width=\linewidth]{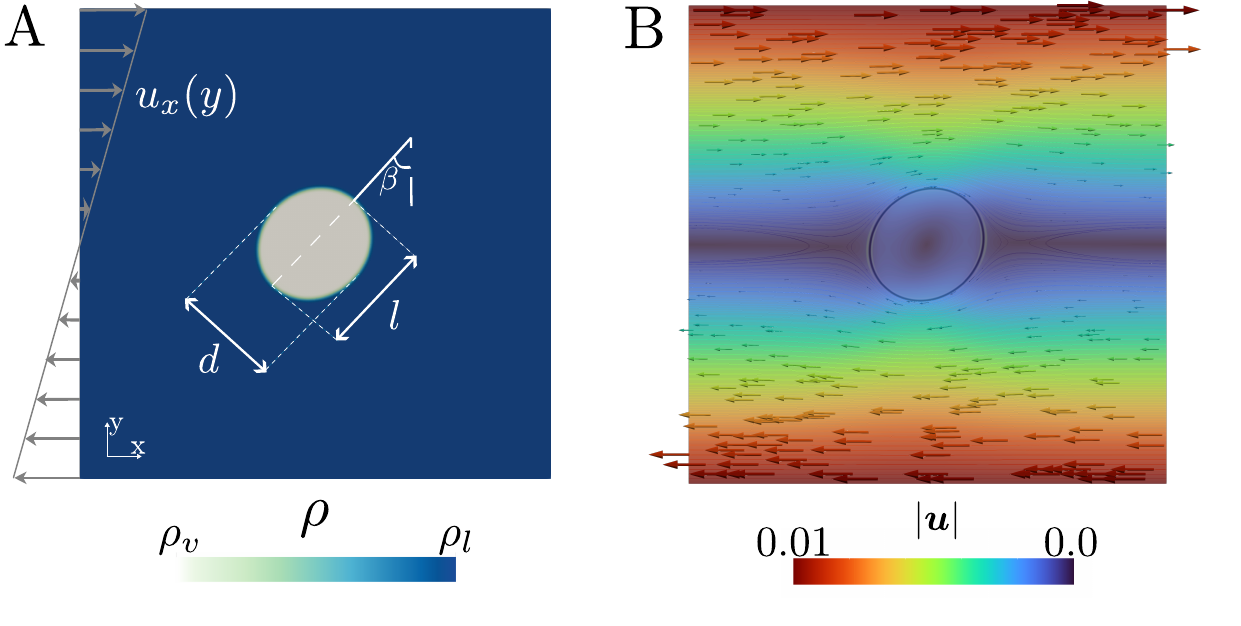}
    \caption{(A) Density contours at equilibrium, illustrating the deformation of a single bubble in shear flow. The bubble deforms into an elliptical shape, characterized by a major axis diameter $l$, a minor axis diameter $d$, and an angle of inclination $\beta$. (B) Contour of velocity magnitude. Color arrows are scaled by local velocity magnitude and indicate the flow direction. The black solid line represents the vapor-liquid interface.}
    \label{fig:shearflow_illustration}
\end{figure}

Fig.~\ref{fig:shearflow_illustration} shows our computational domain (a square of side 1) and some quantities of interest. The initial velocity is $u_x(\boldsymbol{x},0)=0.01(2y-1)$ and $u_y(\boldsymbol{x},0)=0$, which generates a shear flow in the horizontal direction that progressively deforms the vapor bubble. The vapor bubble has an initial radius $R=0.12$ and is placed at the center of the domain. The construction of the initial conditions of density and concentration fields follows the same procedures described in Section~\ref{LP}. At the top and bottom walls, we impose Dirichlet boundary conditions that are consistent with the initial conditions for density, velocity, and surfactant concentration. We use periodic boundary conditions on the right and left sides. We set the $\text{Ca}=1/256$, $\text{Re}=100$ and $\text{Pe}=145$. 

\begin{figure}
    \centering
    \includegraphics[width=\linewidth]{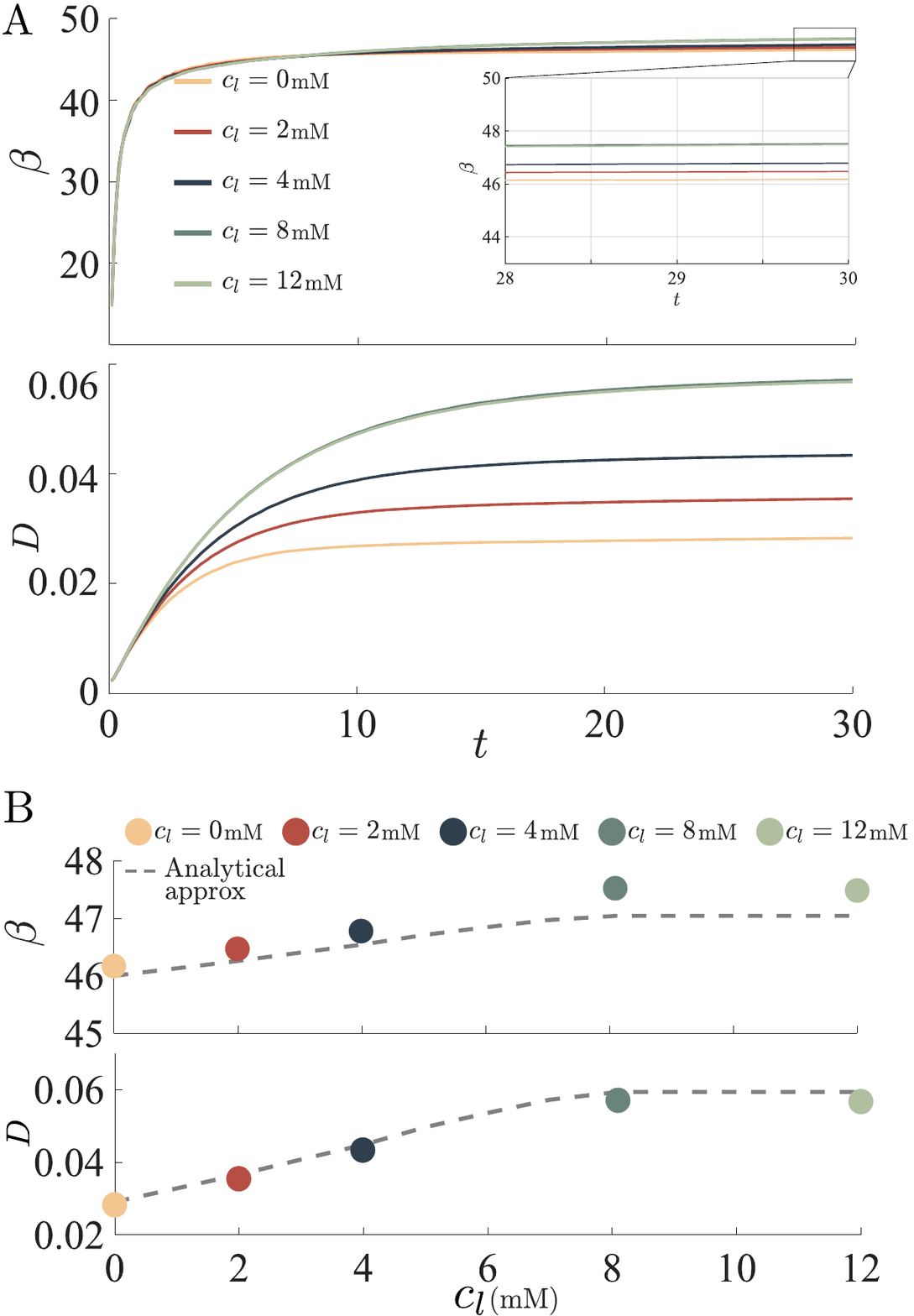}
    \caption{(A) Top: time evolution of the inclination angle $\beta$ of the bubble in shear flow for different values of $c_l$; Bottom: Time evolution of $D$ for different values of $c_l$. (B) Comparison of $\beta$ (top) and $D$ (bottom) using theoretical estimates (dashed line) and simulation results (circular markers).}
    \label{fig:shearflow_results}
\end{figure}

When the simulation starts, the bubble begins to deform from its initial circular shape. After sufficient time, the bubble reaches an equilibrium shape. We show the time evolution of the deformation by reporting the angle of inclination $\beta$ and the parameter $D=(l-d)/(l+d)$ for different values of the surfactant concentration in the bulk liquid; see  Fig.~\ref{fig:shearflow_results}A. The angle of inclination $\beta$ increases mildly with the surfactant concentration. The parameter $D$ is much more sensitive to the surfactant concentration. For $c_l=0$ mM, the strong surface tension effects maintain the bubble's shape close to its original spherical form. The minor deformation induced by the shear flow reaches equilibrium within a short period ($t<10$). As the surfactant concentration grows, the equilibrium shape deviates more markedly from spherical, as quantified by $D$. Given that the CMC for the SDS solution is around 8 mM, the surface tension remains constant for values $c_l$ that exceed 8 mM. Therefore, the $D$ curves for concentrations higher than the CMC practically overlap. 

To gain further insight, we quantitatively compare the equilibrium values of $\beta$ and $D$ obtained from the simulations with theoretical estimates; see Fig.~\ref{fig:shearflow_results}B. To perform this comparison, we define the capillary number for shear flow
\begin{equation}
    \text{Ca}_s=\frac{R G \bar{\mu}}{\sigma}, 
\end{equation}
where $G=|u_x(1)-u_x(0)|$ is the shear rate for the unit domain. For different $c_l$, we use the corresponding $\sigma$ reported in our simulations in Fig.~\ref{fig:laplace_pressure_tot}. The resultant $\text{Ca}_s\in[0.0286, 0.0591]$ for varying $c_l$. Under the condition that $\text{Ca}_s\ll1$, the theoretical estimates for $\beta$ and $D$ are \cite{Rust2002-lh}
\begin{equation}
    \begin{split}
        &\beta = \frac{\pi}{4} + 0.6\text{Ca}_s,\\
        &D = \text{Ca}_s.
    \end{split}
\end{equation}
Fig.~\ref{fig:shearflow_results} shows that our computational results yield a maximum relative difference of 1\% and 3\% in $\beta$ and $D$, respectively. 

\subsection{Effect of surfactant on bubble coalescence and condensation}

\begin{figure}
    \centering
    \includegraphics[width=\linewidth]{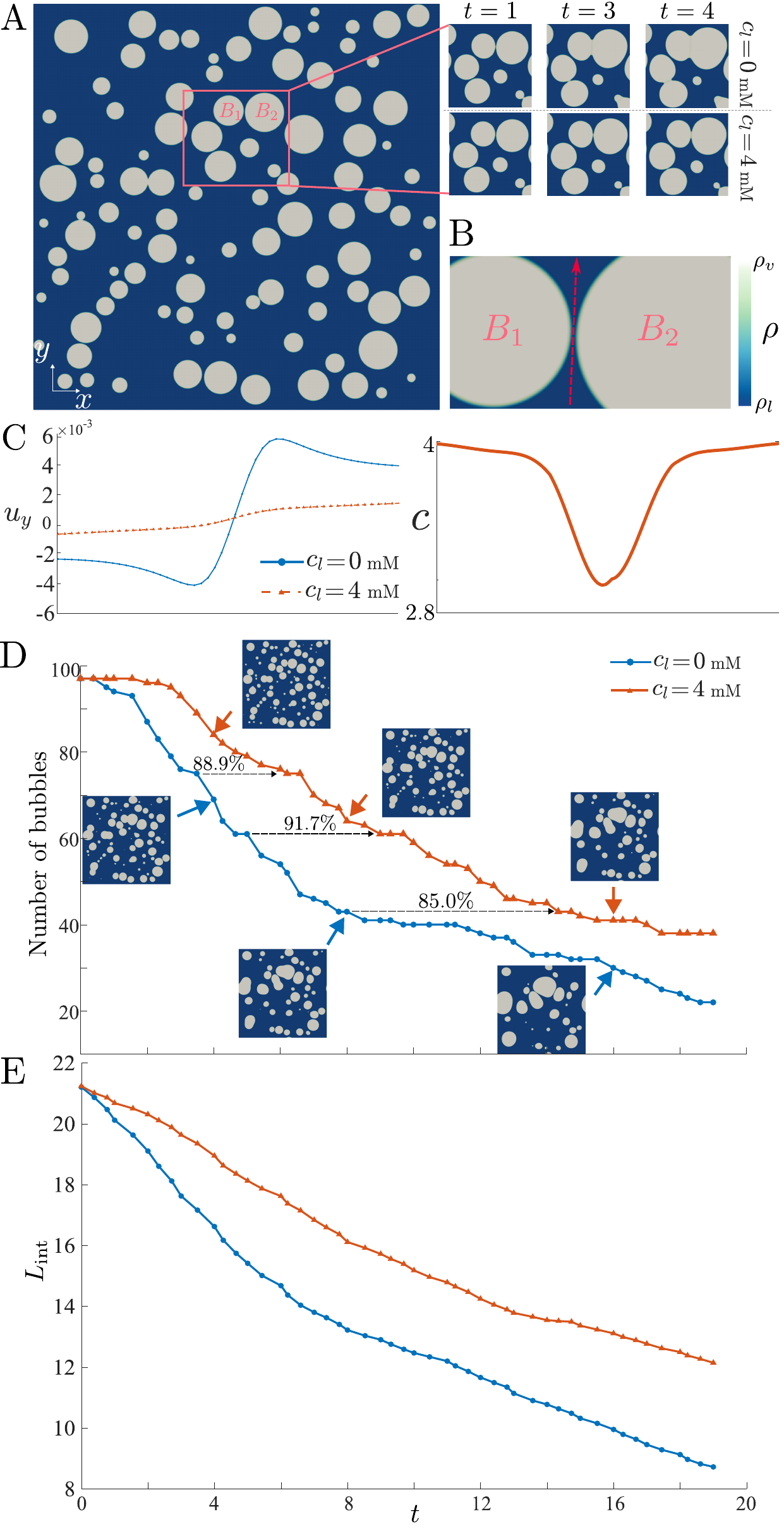}
    \caption{Bubble coalescence in pure water ($c_l=0$ mM) and water-surfactant mixture ($c_l=4$ mM). We use $2500\times2500$ elements in the simulation, and $\text{Ca}=1/1250$, $\text{Re}=100$, $\text{Pe}=700$. (A) We create 100 vapor bubbles at the initial state (left), and focus on a certain area to observe the coalescence of the bubbles at $c_l=0$ mM and $c_l=4$ mM (right). (B) We draw a line across the thin liquid film between $B_1$ and $B_2$, and plot the velocity along the line at $t=1$ in (C) (left). The velocity is measured from bottom to top. Positive values mean upward velocity. The $c$ distribution in the liquid film is drawn in (C) (right). (D) Time evolution of the number of bubbles in the domain. (E) Time evolution of the total length of interface in the domain. }
    \label{fig:multiBubb_tot_new}
\end{figure}

Here, we consider a large number of vapor bubbles in a pool of liquid under conditions close to equilibrium. The number of vapor bubbles will decrease over time via two mechanisms: bubble coalescence and condensation. Coalescence occurs when neighboring bubbles merge due to liquid-film drainage between them. Condensation is stronger in bubbles smaller than a certain threshold in the near-equilibrium conditions studied here. For small bubbles, the equilibrium pressure is high due to their large curvature---but for pure vapor bubbles, the pressure cannot grow much higher than vapor pressure, which leads to bubble condensation. 

The presence of a surfactant changes how bubbles coalesce and condense. First, it slows coalescence. When two bubbles are close to each other, the surfactant molecules distribute unevenly along the bubble interfaces due to flow and bubble deformation. The spatial variation creates local surface-tension gradients that drive Marangoni stresses opposing interface deformation \cite{Takagi2011-hc}. Combined with the surfactant’s overall reduction of mean surface tension, these Marangoni stresses slow the liquid film drainage between neighboring bubbles, thereby reducing coalescence rates \cite{Karakashev2015-ky,Lu2019-nr}. Second, the presence of a surfactant reduces the mean surface tension, thereby reducing  Laplace pressure and making small bubbles less prone to condensation.

To quantitatively assess the surfactant effect on these two mechanisms, we conducted a simulation with 100 near-equilibrium vapor bubbles within a square domain {$\Omega=(0,1)\times(0,1)$}; see Fig.~\ref{fig:multiBubb_tot_new}A. We used periodic boundary conditions. To better understand the mechanisms that inhibit bubble coalescence and condensation, we focus on two specific bubbles, namely $B_1$ and $B_2$ (Fig.~\ref{fig:multiBubb_tot_new}A-B). 

At $c_l=0$ mM, the two proximal bubbles $B_1$ and $B_2$ come into contact before $t=3$, whereas, with the surfactant, this occurs at  $t\approx4$. In pure water, by $t=4$, $B_1$ and $B_2$  have merged into a single vapor region, which eventually evolves to a spherical shape due to surface tension. To better understand the details of bubble coalescence, we draw a line in the liquid film between $B_1$ and $B_2$ at $t=1$; see Fig.~\ref{fig:multiBubb_tot_new}B. We plot the $y$-velocity along this line in Fig.~\ref{fig:multiBubb_tot_new}C (left). A positive value indicates upward flow, while a negative value indicates downward flow. The draining speed of liquid in the thin film is significantly lower with surfactant. To unveil the reason for the velocity reduction, we plot $c$ along the same line in Fig.~\ref{fig:multiBubb_tot_new}C (right). The surfactant concentration is much lower in the middle of the liquid film than in the surrounding bulk liquid. Such depletion of surfactant at the film center agrees qualitatively with the results in \cite{Chesters2000-rd}. The uneven surfactant distribution generates local surface tension gradients that drive liquid toward regions with higher surface tension due to the Marangoni effect, which opposes the drainage of the liquid film between the bubbles and delays coalescence.

We use topology discrimination to count bubbles in scenarios with $c_l=0$ mM and $c_l=4$ mM. Initial contact among some bubbles results in a count of 97 at $t=0$ (Fig.~\ref{fig:multiBubb_tot_new}D). Under the influence of coalescence and condensation, the bubbles in the water-surfactant mixture exhibit remarkable stability from $t=0$ to $t=3$, with the bubble number remaining almost constant. In the presence of surfactant, the rate of reduction in the number of bubbles is over 85$\%$ lower throughout most of the simulation period, highlighting the substantial influence of the surfactant on bubble stability.

The time evolution of the total length of the interface $L_\text{int}$ in the domain is shown in Fig.~\ref{fig:multiBubb_tot_new}E. The interface is identified by the iso-contour $\rho=\rho_\text{mid}\equiv(\rho_v+\rho_l)/2$. We observe that the interfacial length decreases faster in pure water than in the surfactant-laden cases. This behavior indicates stronger curvature relaxation and more rapid bubble-bubble merging in the absence of surfactant.

\begin{figure}
    \centering
    \includegraphics[width=\linewidth]{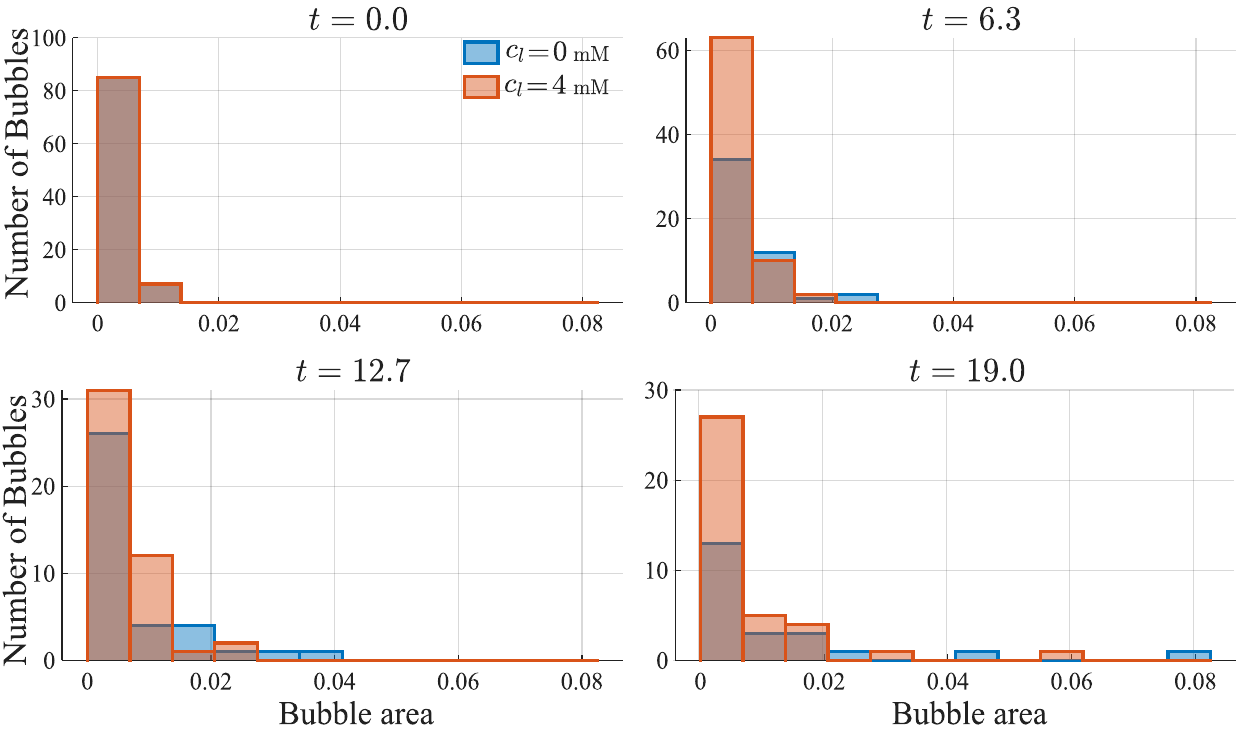}
    \caption{Histograms of bubble area at four time instances for cases $c_l=0$ mM and $c_l=4$ mM.}
    \label{fig:multiBubb_area}
\end{figure}
Fig.~\ref{fig:multiBubb_area} shows the histogram of bubbles' area at four time instances. The histograms illustrate that, as the total number of bubbles decreases, small bubbles persist longer and remain more numerous in the presence of surfactant, whereas without surfactant, they more readily coalesce into larger bubbles.

\subsection{Coalescence of two bubbles in three dimensions}

\begin{figure}
    \centering
    \includegraphics[width=\linewidth]{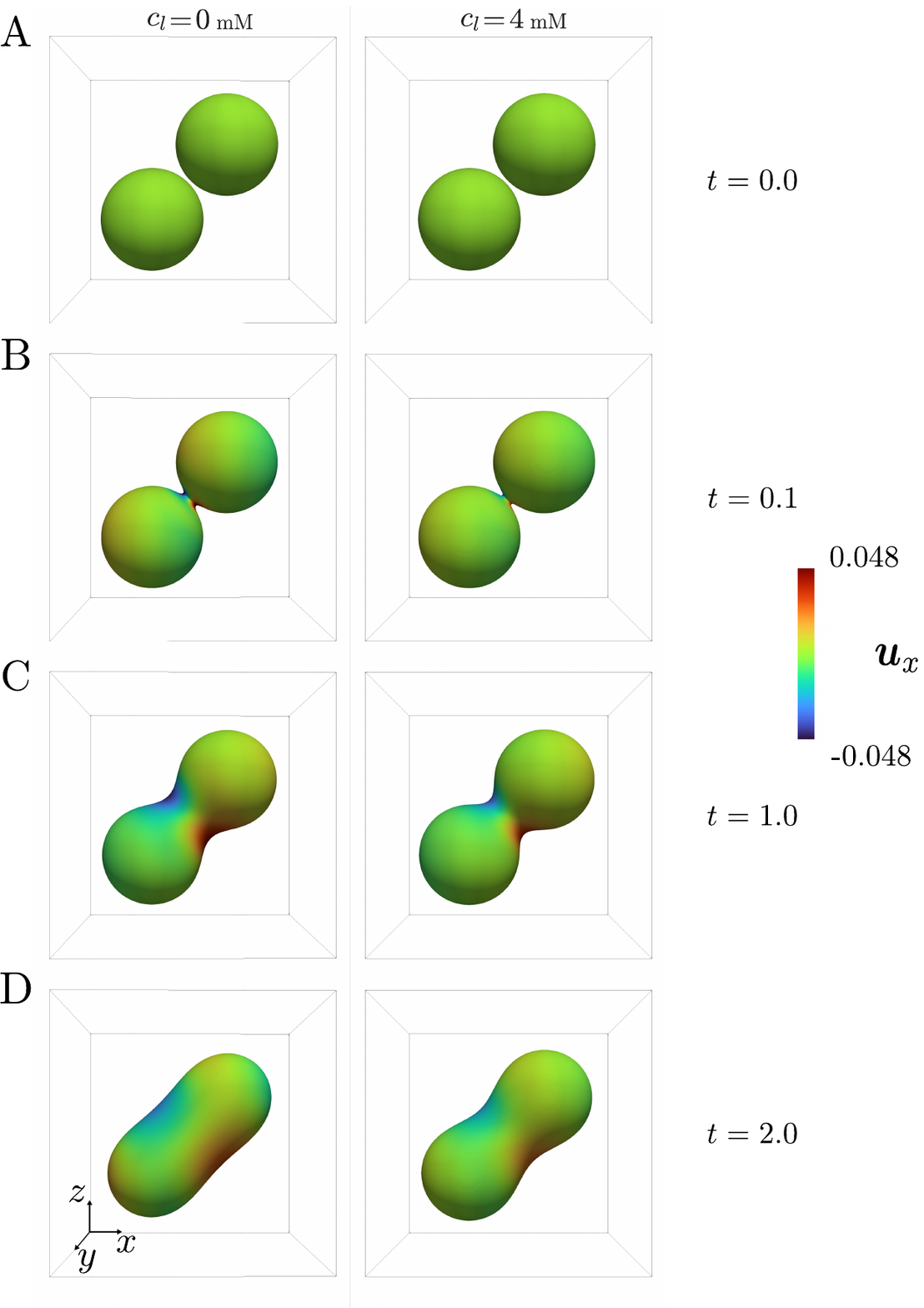}
    \caption{Time evolution of the density iso-surfaces of two adjacent bubbles in a 3D computational domain. The color scale on the isosurfaces represents the $x$-direction velocity. (A) Initial condition. (B) Early-contact phase. (C) and (D) As time evolves, the merged bubble gradually relaxes from a dumbbell-shaped morphology toward a more spherical shape.}
    \label{fig:3D}
\end{figure}
To demonstrate that the present framework can be extended to three-dimensional and more complex configurations, we conducted a 3D simulation showing the coalescence of two bubbles. Fig.~\ref{fig:3D} shows the iso-surfaces of the bubbles at $\rho=\rho_\text{mid}$. The domain is the box $\Omega=(0,1)\times(0,1)\times(0,1)$ which we discretized with $128\times128\times128$ elements. We set $Ca=1/128$, $Re=128$, and $Pe=72$. The centers of the two bubbles are set to be $C_1=(0.34, 0.5, 0.34)$ and $C_2=(0.66, 0.5, 0.66)$, respectively, with same radius $R=0.22$. For each grid point $\boldsymbol{x}$, the signed distance to each bubble interface is computed as
\begin{equation}
    \mathcal{X}_1=|\boldsymbol{x}-C_1|-R, 
    \quad \mathcal{X}_2=|\boldsymbol{x}-C_2|-R.
\end{equation}
The combined signed-distance field is defined as
\begin{equation}
    \mathcal{X} = \min(\mathcal{X}_1, \mathcal{X}_2),
\end{equation}
The density field is then initialized as
\begin{equation}
    \rho(\boldsymbol{x})=\frac{\rho_v+\rho_l}{2} + \frac{\rho_l-\rho_v}{2}\tanh{\frac{\mathcal{X}}{\text{Ca}}},
\end{equation}
and the initial condition for the surfactant concentration is $c=g c_l$. We simulate two cases, $c_l=0$ mM and $c_l=4$ mM. At the initial contact phase $t=0.1$, as the bubbles coalesce faster in the absence of surfactant, the width of the neck connecting the two bubbles is wider without surfactant than  with surfactant. Later, the expansion speed of the connecting neck is also higher without surfactant. At times $t=1.0$ and $2.0$, the merged bubble relaxes toward a spherical shape much faster for $c_l=0$ mM than for $c_l=4$ mM. These results further demonstrate that the present model can be extended to 3D configurations and more complex scenarios. Further validation may be pursued through comparison with recent studies in multiphase flow \cite{Dong2024-om, Reese2024-la, Zhang2025-gh} under more extreme situations.

\section{Conclusion}

We proposed a mathematical model to understand the effect of surfactants on liquid-vapor phase transformations. The model is based on the NSK equations, which have been used to model cavitation under the framework of DVS. One strength of the NSK equations is that they do not require the use of mass-transfer functions, which involve parameters that depend on the flow conditions and need frequent recalibration. Thus, the NSK equations constitute an ideal framework for modeling liquid-vapor phase change in non-ideal conditions, including the presence of surfactants, particulate matter, or other water contaminants.

We validated the model’s ability to reproduce surfactant-mediated reduction in surface tension during phase transformations by comparing the computed interfacial tensions with experimental data for SDS over the full concentration range, including sub-CMC concentrations ($0\sim8$ mM) and super-CMC concentrations ($8\sim12$ mM). In the context of capillary oscillation, surface tension changes were also confirmed by comparing the oscillation frequency of the interface with a theoretical estimate. We also demonstrated the model’s capability to capture key surfactant-induced phenomena, including enhanced bubble deformability in shearing flow, inhibition of bubble coalescence, and delay of bubble condensation. A three-dimensional case was performed to illustrate the extensibility of the present framework to more complex situations.

Overall, this work establishes a robust and general framework for predicting surfactant effects on liquid-vapor phase transformation in a thermodynamically consistent manner. The formulation is sufficiently general to be applied to more complex flow configurations and interfacial topologies. It also provides a natural basis for extension to non-isothermal settings, where heat transfer and temperature-dependent surface-tension variations can be incorporated.


\section*{Acknowledgements}
This work was partially supported by the U.S. Office of Naval Research (PO Dr. Yin Lu (Julie) Young; Award No. N000142512096),
This work uses the Bridges-2 system at the Pittsburgh Supercomputing Center (PSC) through allocation no. MCH220014 from the Advanced Cyberinfrastructure Coordination Ecosystem: Services and Support (ACCESS) program, which is supported by the National Science Foundation, grant nos. 2138259, 2138286,
2138307, 2137603, and 2138296.

\nocite{*}
\bibliography{paperpile}

\end{document}